\documentclass[twocolumn,showpacs,amsfonts,aps,prc,nofootinbib,floatfix]
{revtex4}

\usepackage{amsmath}
\usepackage{bm}
\usepackage{graphicx}

\usepackage{epsfig}
\newcommand{\beq}{\begin{equation}}
\newcommand{\eeq}{\end{equation}}
\newcommand{\bea}{\vspace{0.25cm}\begin{eqnarray}}
\newcommand{\eea}{\end{eqnarray}}

\newcommand{\ro}{\mbox{{\boldmath
$\rho$}}}

\newcommand{\qb}{\mbox{{\bf
q}}}

\newcommand{\bb}{{{\bf b}}}

\def\lsim{\mathrel{\rlap{\lower4pt\hbox{\hskip1pt$\sim$}}
    \raise1pt\hbox{$<$}}}         
\def\gsim{\mathrel{\rlap{\lower4pt\hbox{\hskip1pt$\sim$}}
    \raise1pt\hbox{$>$}}}         

\begin{document}


\title{
Medium-modification of photon-tagged jets in $AA$ 
collisions
}
\date{\today}

\author{B.G.~Zakharov$^{a,b}$}

\affiliation{a) Center for Interdisciplinary Studies in Physics and Related Areas, Guizhou University of Finance and Economics, Guiyang 550025, China\\
b) L.D.~Landau Institute for Theoretical Physics,
        GSP-1, 117940, Kosygina Str. 2, 117334 Moscow, Russia}

\begin{abstract}
We study nuclear modification of the photon-tagged jets 
in $AA$ collisions within the jet quenching scheme
based on the light-cone path integral approach to the induced gluon emission.
The calculations are performed for running coupling.
Collisional energy loss is treated as a perturbation
to the radiative mechanism.
We obtain a reasonable agreement with the recent data from
the STAR Collaboration on the mid-rapidity nuclear modification factor $I_{AA}$ for 
Au+Au collisions at $\sqrt{s}=200$ GeV
for parametrization of running $\alpha_s$ 
consistent with that 
necessary for description of the data on 
suppression of the high-$p_T$ spectra.

\end{abstract}
%

\maketitle

\section{Introduction}

Results from RHIC and LHC on heavy-ion collisions
give strong evidence for production
of a deconfined quark gluon plasma (QGP).
One of the main signature of the QGP formation in $AA$ collisions
is the discovery at RHIC and LHC of the extremely strong suppression
of the high-$p_T$ hadron spectra. 
It is commonly believed that this suppression
is a consequence of the jet modification
(jet quenching) due to the final state interaction with the QGP produced
in the initial stage of $AA$ collisions.
The jet quenching is caused by the radiative 
\cite{BDMPS,LCPI,W1,GLV1,AMY} 
and collisional \cite{Bjorken1} 
energy loss of fast partons in the QGP.
The RHIC and LHC data on the nuclear modification factor 
$R_{AA}$, characterizing the suppression of the high-$p_T$ spectra,
can be reasonably described by the radiative and collisional 
parton energy loss in the QGP
with dominant contribution from the
radiative mechanism due to the induced gluon emission.
A consistent analysis of the jet quenching phenomenon requires
understanding multiple gluon emission.
The available approaches to the induced gluon emission 
\cite{BDMPS,LCPI,GLV1,AMY} deal with one gluon emission.
At the one gluon level 
in the light-cone path integral (LCPI) \cite{LCPI} approach
the spectrum of gluon emission by a quark
may be expressed via the retarded Green function of a two dimensional
Schr\"odinger equation, in which the longitudinal coordinates 
(along the initial quark momentum) plays the role of time,
and the imaginary potential is proportional the cross section  
of interaction of the three-body $q\bar{q}$-system with the medium
constituent. The diagram technique developed in Refs.~\cite{LCPI}
allows one to go beyond the one gluon level.
However, already for the double gluon emission, even in a crude 
oscillator approximation \cite{Z_OA,AZ} (when the potential 
is approximated by a quadratic form), calculations become extremely 
complicated \cite{Arnold_2g}. And up to now there are no
phenomenological schemes for the jet quenching analyses 
that treat accurately the double gluon emission.
Anyway the double gluon level is insufficient for analyses of 
the jet quenching data from RHIC and LHC.
In the presently available analyses of the nuclear modification
factor $R_{AA}$ the effect of multiple gluon emission is usually
accounted for in the approximation of independent 
gluon emission \cite{RAA_BDMS,GLV2,Eskola,RAA08},
similar to the Landau method for multiple soft photon emission in QED. 
This approximation does not account
for the effect of the gluon cascading that may be important 
for the medium-modified fragmentation functions (FFs)
in the soft region $z\ll 1$.
Nevertheless, this approximations seems reasonable to calculate the nuclear
modification factor $R_{AA}$. Because it  depends mostly on the form
of the medium-modified FFs  for parton$\to$hadron
transitions 
in the region of intermediate and large  
$z$, where the main effect of multiple gluon
emission is the Sudakov suppression which should not be sensitive
to the details of the in-medium parton cascading at $z\ll 1$.

Since at small $z$ the approximation of independent gluon emission 
becomes questionable, 
it is of course highly desirable
to perform comparison 
of the theoretical predictions obtained in the approximation
of independent 
gluon emission \cite{RAA_BDMS,GLV2,Eskola,RAA08}
with 
experimental observables
that are sensitive to the form of the medium-modified FFs
in a broad range of $z$.   
Experimentally the information about the medium jet modification
in a broad range of $z$
can be obtained from measurement of the photon-tagged jet FFs 
in $\gamma$+jet events
\cite{Wang1}. 
The medium effects in jet fragmentation
in $\gamma$+jet events
are characterized by the nuclear modification factor $I_{AA}$, 
defined as the ratio
\beq
I_{AA}=\frac{D^{AA}_{h}(z)}{D^{pp}_{h}(z)}\,,
\label{eq:10}
\eeq
where $D^{AA}_{h}(z)$ and $D^{pp}_{h}(z)$  are the $\gamma$-triggered jet FFs 
for $AA$ and $pp$ collisions, respectively.
The mid-rapidity factor $I_{AA}$ has been recently measured by 
the STAR Collaboration 
\cite{STAR1} at RHIC
for central Au+Au collisions at $\sqrt{s}=200$ GeV 
in a broad range of $z$ for the photon energies $12< E_{T}^{\gamma}<20$ GeV.

In the present paper we perform a comparison of the STAR data \cite{STAR1}
with the theoretical
predictions for $I_{AA}$ obtained within the model of jet quenching that we 
developed in Ref.~\cite{RAA08}, and 
previously
used in Refs.~\cite{RAA11,RAA12,RAA13} for successful description of the 
data on 
the nuclear modification factor $R_{AA}$.
The scheme is based on the LCPI approach \cite{LCPI} to the induced gluon
emission. The method allows one to treat accurately the 
Landau-Pomeranchuk-Migdal (LPM) effect and the finite-size 
effects.
The calculations are performed for running coupling.
We perform numerical calculations 
beyond the oscillator approximation when parton
multiple scattering in the medium can be described 
in terms the well known transport coefficient $\hat{q}$ \cite{BDMPS}.
The model also includes the contribution of the collisional energy loss.

The plan of the paper is as follows. In Sec.~2 
we review our theoretical framework.
In the first subsection
we 
discuss our model for the in-medium FFs.
In the second subsection we discuss the model of the QGP fireball
used in our numerical calculations.
In the last subsection we discuss calculations of the nuclear
modification factor $I_{AA}$.
In Sec.~3 we present comparison of our numerical results
with the STAR data \cite{STAR1} on 
the nuclear modification factor $I_{AA}$.
Sec.~4 summarizes our work.

\section{The theoretical framework}
In this section, we review the main aspects
of our theoretical framework for
calculation
of the nuclear modification factor $I_{AA}$ that
characterizes the jet medium modification in $AA$ collisions.

\subsection{Medium-modified FFs}
As was said in Introduction, currently 
the first principle analysis of  the jet medium  modification 
in $AA$ collisions is impossible. Our treatment of
the medium modified FFs ${D}_{h/i}^{m}$ for parton$\to$ hadron transitions
for $AA$ collisions
is similar to
that developed in our analysis 
\cite{RAA08}
of the nuclear modification factor $R_{AA}$. It is
based on the LCPI approach \cite{LCPI} to the induced gluon emission
from fast partons in the QGP.
For reader's convenience, and since some of the details
have been omitted in our concise paper \cite{RAA08}, 
in this subsection we discuss 
the important points of our model for calculation
of the medium modified FFs.

We assume that that the parton$\to$hadron 
transition consists of the three stages: the DGLAP cascading, the induced
gluon emission stage in the QGP, and parton hadronization outside the QGP.
For a given jet trajectory in the fireball the FF ${D}_{h/i}^{m}$ reads 
\beq
 {D}_{h/i}^{m}(z,Q)= \int_z^1\frac{dz'}{z'}
D_{h/j}(z/z',Q_{0})D_{j/i}^{m}(z',Q)\,,
\label{eq:20}
\eeq
where $D_{h/j}$ describes parton hadronization
outside the QGP, and ${D}_{j/i}^{m}$ corresponds 
to transition of the initial hard parton $i$ to
the parton $j$ escaping from the QGP. 
The partonic FF ${D}_{j/i}^{m}$
includes the parton evolution in the DGLAP stage and 
medium modification in the QGP. We write it as a convolution
\bea
 {D}_{j/i}^{m}(z,Q)= 
\int_z^1\frac{dz'}{z'}
D_{j/k}^{in}(z/z',E_k)
\nonumber\\
\times D_{k/i}^{DGLAP}(z',Q_{0},Q)\,,
\label{eq:21}
\eea
where $E_k=z'Q$. 
The FF $D_{k/i}^{DGLAP}(z,Q_0,Q)$  describes the first
DGLAP stage for parton$\to$parton transition 
in the parton cascading 
from the initial parton virtuality 
$Q$ to a small virtuality scale $Q_0$, where the DGLAP
cascade is stopped. The  FF $D_{j/k}^{in}(z,E_k)$
corresponds to the in-medium
parton$\to$parton transition in the QGP fireball.
It depends on the energy of the parton $k$.

In the absence of the medium the $D_{j/k}^{in}$ in Eq.~(\ref{eq:21})
reduces to the unit operator 
$\delta_{jk}\delta(z-1)$, and
$D_{j/i}^{m}(z,Q)$ becomes equal to 
$D_{j/i}^{DGLAP}(z,Q_{0},Q)$.
In this case (\ref{eq:20}) reduces
to its vacuum counterpart corresponding to FF for $pp$ collisions 
\bea
 {D}_{h/i}(z,Q)= \int_z^1\frac{dz'}{z'}
D_{h/j}(z/z',Q_{0})
\nonumber\\
\times D_{j/i}^{DGLAP}(z',Q_0,Q)\,.
\label{eq:30}
\eea
As in Ref.~\cite{RAA08} we take $Q_0=2$ GeV
for the FFs ${D}_{h/j}(z,Q_0)$ in Eq.~(\ref{eq:20}),   
describing parton$\to$hadron
transition outside the QGP (and in Eq.~(\ref{eq:30}) for $pp$ collisions).
For these FFs we use the KKP \cite{KKP} parametrization.
The DGLAP FFs 
$D_{k/i}^{DGLAP}(z,Q_{0},Q)$   have been computed with the help of 
the PYTHIA event generator \cite{PYTHIA}. It was used to create
a grid of values for $D_{k/i}^{DGLAP}(z,Q_{0},Q)$   in the $z-Q$ plane. 
Our method for calculation of the FFs for $pp$ collisions in the 
form of the convolution of the KKP FF at $Q_0=2$ GeV and the DGLAP
FFs guarantees that in the limit of the vanishing induced radiation
the medium-modified FFs given by Eqs.~(\ref{eq:20}), (\ref{eq:21})
exactly reduce to the $pp$ FFs (\ref{eq:30}). We checked that quantitatively
our formula (\ref{eq:30}) reproduces reasonably well the $Q$-dependence 
of the KKP FFs \cite{KKP}. Nevertheless, to avoid the effect of 
a possible difference between the KKP FFs and the FFs given
by Eq.~(\ref{eq:30}) on predictions for $I_{AA}$, 
the use of the form (\ref{eq:30}) is clearly preferred for numerical 
calculations of the $I_{AA}$.

The form given by Eqs.~(\ref{eq:20}), (\ref{eq:21})  assumes that the 
DGLAP and the induced gluon
emission stages are approximately ordered in time. 
This picture seems to be reasonable for initial parton energies 
$\lsim 100$ GeV, because in this region the typical formation time 
for emission of the first most energetic gluon in the DGLAP cascade
turns out to be relatively small.
The gluon formation length 
emitted by a fast quark in vacuum is approximately 
\beq
L_{f}(x,k_T)\sim
2E_q x(1-x)/(k_T^2+\epsilon^2)\,,
\label{eq:40}
\eeq
 where $x$ is the gluon fractional momentum, 
and $\epsilon^2=m_q^2x^2+m_g^2(1-x)$. From Eq.~(\ref{eq:40})
using the vacuum spectrum of the gluon emission from a quark
\beq
\frac{dN}{dk_{T}^{2}dx}=\frac{C_{F}\alpha_{s}(k_{T}^{2})}{\pi x}
\left(1-x+x^{2}/2)\right)\frac{k_{T}^{2}}{(k_{T}^{2}+\epsilon^{2})^{2}}\,
\label{eq:41}
\eeq
we obtained that for $E\lsim 100$ GeV
the typical gluon formation length $\bar{L}_f\sim 0.3-0.5$ fm
(we take $m_g=Q_0$ since we are interested in the typical length for gluons
with virtuality $Q\gsim Q_0$).
This says that the hardest gluon emission in 
the DGLAP cascade typically occurs before formation of the equilibrated QGP,
which is expected at the proper time $\tau_0\sim 0.5$ fm. 
The above estimate for the typical time of the DGLAP stage 
agrees with qualitative $L$-dependence 
of the fast parton virtuality $Q(L)\sim \sqrt{Q/L}$ (which can be 
obtained from the uncertainty relation $\Delta E\Delta t\sim 1$).
This says that for $L\sim \tau_0\sim 0.5$ fm for the initial partons with 
$E\sim 10-50$ GeV we have $Q(L)\sim 2-4$ GeV.
For this reason the scale $Q_0\sim 2$ GeV, that 
is reasonable for the lower end of the DGLAP evolution,
at the same time translates to the longitudinal scale
that agrees qualitatively with the QGP production time where
the induced gluon emission comes into play. 
Note however, that our numerical calculations show that the results 
are practically insensitive to the value of $Q_0$.
For this reason it does not make sense to try to find more appropriate
value of $Q_0$ which better corresponds to the space-time picture 
of the QGP production and transition of the DGLAP stage to the stage 
of the induced gluon emission in the QGP. 
Anyway, in the DGLAP cascade the time of parton
splitting can be only estimated very roughly. 
One remark about the arguments of the FFs in Eqs.~(\ref{eq:20}), (\ref{eq:21}) 
is in order. 
For the DGLAP and the hadronization stages the FFs in 
Eqs.~(\ref{eq:20}), (\ref{eq:21}) 
are written as functions of the parton virtualities. But the in-medium 
FF $D_{j/k}^{in}$ is a function of the parton energy. It is because
we evaluate the induced gluon spectrum within the old fashioned perturbation
theory in the coordinate representation in which particles 
are not characterized by virtuality. 
In this formulation the virtuality may be estimated from the 
length scale $L$ and 
the parton energy $E$  with the help the uncertainty relation that gives $Q\sim \sqrt{E/L}$ which, of course, matches the above formula for $Q(L)$.

We are fully aware that the picture with the time ordering of the DGLAP
and the induced gluon emission stages 
may be questionable
for the DGLAP
gluon emission with sufficiently small transverse momenta 
(about the Debye mass of the QGP) when the vacuum gluon formation length 
becomes as large as the typical formation length for the induced
gluon emission. Note that just because of the interference of the vacuum
gluon emission with the induced one the induced gluon spectrum
vanishes for zero medium size. In the form (\ref{eq:20}), (\ref{eq:21})
these interference effects are assigned to the in-medium FF $D_{j/k}^{in}$.
In the absence of a consistent approach to the in-medium parton cascading
it is difficult to estimate the theoretical uncertainties from the use of
the representation given by Eqs.~(\ref{eq:20}), (\ref{eq:21}). 
It is worth noting that a formal 
interchange in Eq.~(\ref{eq:21}) of the DGLAP and in-medium FFs
practically does not change the results.

We calculate the in-medium FFs $D_{j/k}^{in}(z,E_k)$ in the approximation
of independent induced gluon emission \cite{RAA_BDMS} using for
the one gluon emission distribution the induced gluon spectrum
in the form obtained in Ref.~\cite{Z04_RAA} within the LCPI approach
\cite{LCPI}. The form 
of Ref.~\cite{Z04_RAA} does not 
require calculation of the singular Green's function as in the original
representation of the spectrum of Refs. \cite{LCPI}. The method of 
Ref.~\cite{Z04_RAA}
reduces calculation of the gluon
spectrum to solving a two-dimensional Schr\"odinger equation
in backward time direction
with a smooth boundary condition. It is convenient for
accurate numerical calculations beyond the oscillator approximation.
For the reader's convenience in
 Appendix A we give the necessary formulas.

In the approximation of independent gluon emission the quark fractional 
energy loss distribution in $\xi=\Delta E/E$ can be written as \cite{RAA_BDMS}
(hereafter, for notational simplicity, we omit argument $E$)
\beq
\hspace{-.1cm}W(\xi)\!=\!W_0\sum_{n=1}^{\infty}\frac{1}{n!}\left[
\prod_{i=1}^{n}\int_0^1 dx_{i}
\frac{dP}{dx_i}
\right]\delta\left(\xi-\sum_{i=1}^{n}x_{i}\right)
,
\label{eq:50}
\eeq
where 
\beq
W_0=\exp{\left[-\int_{0}^1 dx \frac{dP}{dx}\right]}
\label{eq:50p}
\eeq
is the no gluon emission probability,
$dP/dx$ is the probability distribution for 
the $q\to gq$ transition with $x=E_g/E_q$.
At $\xi\ll 1$ the main effect of the multiple gluon emission
is the Sudakov suppression. It is well seen from
the approximate calculation of (\ref{eq:50}) 
at the level
of two-gluon emission for the regime
when the relative energy loss
$\Delta E/E=\int_{0}^{1}dx xdP/dx$ 
is much smaller than unity. In this regime,
similarly to the electron energy loss 
\cite{Z_SLAC99}, from (\ref{eq:50})  one can obtain
\bea
W(\xi)\approx
\frac{dP}{d\xi}
\exp\left
[-\int\limits_{\xi}^{1}dx
\frac{dP}{dx}\right]
\left
\{1\right.
\nonumber\\
-\left.\frac{1}{2}
\int\limits_{0}^{\xi}dx_{1}\left[
\frac{dP}{dx_{1}}+\frac{dP}{dx_{2}}-
\frac{dP}{dx_{1}}
\frac{dP}{dx_{2}}
\left(
\frac{dP}{d\xi}
\right)^{-1}\right]
\right\}\,,
\label{eq:51}
\eea
where
$x_1+x_{2}=\xi$.
The exponential Sudakov suppression factor in (\ref{eq:51}) 
reflects a simple fact that
emission of gluons with the fractional momentum bigger than
$\xi$ is forbidden. 

For accurate numerical calculation of the distribution $W(\xi)$ 
given by Eq.~(\ref{eq:50})
it is convenient to rewrite it as a series \cite{GLV2}
\beq
W(\xi)=\sum_{n=1}^{\infty}W_{n}(\xi)\,,
\label{eq:501}
\eeq
where $W_n$ are determined by the recurrence relations
\bea
W_{n+1}(\xi)=\frac{1}{n+1}
\int_{0}^{\xi}dx W_n(\xi-x)\frac{dP}{dx}
\label{eq:51p}
\eea
with
\beq
W_1(\xi)=W_0\frac{dP}{d\xi}\,.
\label{eq:51pp}
\eeq
In numerical calculations we set $dP/dx=0$ at $x<m_g/E_q$
and $1-x<m_q/E_q$.

The expression (\ref{eq:50}) satisfies the relations
\beq
\int_{0}^{\infty}d\xi W(\xi)=1\,,
\label{eq:52}
\eeq
\beq
\int_{0}^{\infty}d\xi \xi W(\xi)=\int_{0}^{1}dx x\frac{dP(x)}{dx}\,.
\label{eq:53}
\eeq
For any value of the ratio $\Delta E/E$, 
the formula (\ref{eq:50}) leads to some leakage 
of the probability and the fractional momentum to the unphysical region of 
$\xi> 1$. The effect is small at $\Delta E/E\ll 1$, but
for the conditions of the jet quenching in $AA$ collisions
when $\Delta E/E$ is not very small, say, $\sim 0.2$ \cite{Z_Ecoll} 
 for light quarks at $E\sim 10-20$ GeV for RHIC conditions,  
the effect may be sizeable.
To ensure the flavor conservation 
\beq
\int_0^1 dz D_{q/q}^{in}(z)=1\,,
\label{eq:54}
\eeq
we define a 
renormalized distribution in the physical region $\xi<1$
\beq 
W_R(\xi)=K_{qq}W(\xi)
\label{eq:55} 
\eeq
with
\beq 
K_{qq}\!=\!\int^{\infty}_{0} d\xi   W(\xi)
\Big{/}\!\int^{1}_{0} d\xi W(\xi)\,.
\label{eq:60}
\eeq
Then this renormalized distribution is used to define the in-medium 
$q\to q$ FF 
\beq
D_{q/q}^{in}(z)\!=\!W_R(1-z))\,.
\label{eq:70}
\eeq

We define the FF for $q\to g$ transition
as
\beq
D_{g/q}^{in}(z)\!=\!K_{gq} dP/dz.
\label{eq:71}
\eeq
We determine the coefficient $K_{gq}$ from the momentum sum rule
\beq
\int_0^1 dz z\left [D_{q/q}^{in}(z)+D_{g/q}^{in}(z)\right]=1\,.
\label{eq:72}
\eeq
In the limit $\Delta E/E\to 0$ $K_{gq}\to 1$. 
Indeed, 
at one gluon emission level the $z$-distribution
for $q\to g$ transition is connected to that for $q\to q$
by interchange of the arguments $z \leftrightarrow 1-z$.
Then after setting the upper limit of $\xi$-integration
in Eqs.~(\ref{eq:52}), (\ref{eq:53}) to $1$, we conclude
that the momentum sum rule (\ref{eq:72}) is satisfied for
$K_{gq}=1$.

For the $g\rightarrow g$ transition we use the following
procedure. In the first step  
we define $D_{g/g}^{in}(z)$ in the region $z>0.5$,  where the Sudakov
suppression is important,  
through the independent gluon emission distribution $W(\xi)$ 
(with $\xi=1-z$) 
given by Eq.~(\ref{eq:50}) using for
$dP/dx$  the $x$-distribution for $g\to gg$ induced transition.
Note that for $g\to gg$ transition due to the $x\leftrightarrow  1-x$ symmetry 
of the function $dP/dx$
we can use $0.5$ for the 
upper limit in $x$-integrations in Eqs.~(\ref{eq:50}), (\ref{eq:50p})
(it means that we view the softest gluon with $x<0.5$ as a radiated gluon). 
In the soft 
region $z<0.5$, where one can expect a strong compensation of
the multiple gluon emission and the Sudakov suppression,
we use simply the one gluon distribution $dP/dx$ (with $x=z$). 
This procedure can 
violate the momentum sum rule
\beq
\int_0^1 dz z D_{g/g}^{in}(z)=1\,,
\label{eq:73}
\eeq
which should be satisfied.
To cure this drawback
we multiply the FF obtained at the first step by a renormalization coefficient
$K_{gg}$ defined from the momentum conservation (\ref{eq:73}).
Note that for the jet path length in the QGP $L\sim 5$ fm and 
the typical jet energy $\sim 15$ GeV, that is relevant
to conditions of the STAR experiment \cite{STAR1}, the necessary 
values of the renormalization coefficients turn out to be not very far from 
unity  ($K_{qq}\approx 1.1$, $K_{gq}\approx 0.8$ and $K_{gg}\approx 0.7$).
This says that we are in a regime when the leakage of the probability to the unphysical region
$\xi>1$ and the violation of the momentum sum rule for the 
distribution (\ref{eq:50}) are relatively small.

Note that in our calculations the process $g\to q\bar{q}$ 
is included into the DGLAP FF in 
Eq.~(\ref{eq:21}),
but
we neglect the induced gluon conversion into $q\bar{q}$ pairs
in calculations of the in-medium FF $D_{j/k}^{in}$.
Calculations within the LCPI formalism \cite{LCPI},
using the formulas given in Appendix A, show that for light 
quarks for the QGP produced in $AA$ collisions for RHIC and LHC conditions 
the probability of the induced $g\to q\bar{q}$ transition
turns out to be relatively small. For conditions of the STAR 
data \cite{STAR1}, when the typical gluon energy $E\sim 15$ GeV and 
the typical path length in the QGP $L\sim 5$ fm, the probability
of the gluon conversion into the $q\bar{q}$ states $\lsim 10$\%.
From the point of view of the nuclear modification factors $I_{AA}$ 
and $R_{AA}$ the effect of this process should
be negligible since the hadronization of the $q\bar{q}$ state 
should be similar to that of a gluon.

The above formulas do not include the effect of the 
the collisional energy loss.
Presently there is no a consistent approach for 
incorporating of the collisional energy loss in jet quenching
calculations on an even footing with the radiative mechanism.
In the present analysis, as in Refs.~\cite{RAA08,RAA11,RAA12,RAA13}, 
we treat the collisional energy loss 
(that is relatively small \cite{Z_Ecoll})
 as a perturbation 
to the radiative mechanism,
and incorporate it 
by a small renormalization of the initial QGP temperature 
for the radiative in-medium FFs $D_{i/k}^{in}$
according to the 
change in the $\Delta E$ due to the collisional energy 
loss (see Ref.~\cite{RAA08} for details).   
We evaluate the collisional energy loss 
using the Bjorken method \cite{Bjorken1}, but
with an accurate
treatment of kinematics of the binary collisions (the details can be found
in Ref.~\cite{Z_Ecoll}).

In calculations of the induced gluon emission distribution
$dP/dx$ we take for parton masses the quasiparticle masses in the 
QGP. We use the quasiparticle 
masses $m_{q}=300$ and $m_{g}=400$ MeV  supported by 
the analysis of the lattice data \cite{LH}. Note that the
results are practically insensitive to the quark mass. 
We use 
the Debye mass $\mu_D$ in the QGP obtained in the lattice 
analysis \cite{Bielefeld_Md}, which gives $\mu_{D}/T$ slowly 
decreasing with $T$  
($\mu_{D}/T\approx 3$ at $T\sim 1.5T_{c}$, $\mu_{D}/T\approx 2.4$ at 
$T\sim 4T_{c}$). However, the sensitivity of the results to the Debye mass
is relatively weak. Because the spectrum $dP/dx$ is mostly 
controlled by the behavior of the dipole cross section (see Eq.~(\ref{eq:a50}))
in the region $\rho\lsim 1/\mu_D$, where it depends on $\mu_D$ 
only logarithmically.

Both for radiative and collisional mechanisms
we use 
the one-loop running $\alpha_{s}$ 
frozen at low momenta at
some value $\alpha_{s}^{fr}$ (see Appendix A). 
The use of the same 
parametrization of running $\alpha_s$ for the radiative and collisional
mechanisms is important for minimizing the theoretical 
uncertainties in the fraction  of the collisional contribution.
The results of the analyses of the low-$x$ structure functions 
\cite{NZ_HERA} and 
of the heavy quark energy loss in vacuum \cite{DKT} show that for 
gluon emission in vacuum for this parametrization  
$\alpha_{s}^{fr}\approx 0.7-0.8$.
However, the thermal effects can suppress the in-medium QCD coupling, 
and we treat $\alpha_{s}^{fr}$  as a free parameter.

\subsection{Model of the QGP fireball}
As in our previous analyses of the nuclear modification
factor $R_{AA}$ \cite{RAA08,RAA11,RAA12,RAA13} we use the ideal gas model of the QGP
with Bjorken's  1+1D expansion \cite{Bjorken}. It gives 
$T_{0}^{3}\tau_{0}=T^{3}\tau$, where $\tau_0$
is the thermalization time of the matter. We take $\tau_{0}=0.5$ fm. 
As in Refs.~\cite{RAA08,RAA11,RAA12,RAA13}, for simplicity, we neglect variation of $T_{0}$ with the transverse coordinates.
We take the medium density 
$\propto \tau$ for $\tau<\tau_{0}$.
This is just an ad hoc prescription to account for the fact that
the QGP production is clearly not an instantaneous process.
Note however, that the effect of the region $\tau<\tau_{0}$
is small. It is due to a strong finite-size suppression of the induced gluon
emission in the regime when the parton path length
is smaller than the gluon formation length \cite{Z_OA,AZ}.

We fix the initial temperature of the plasma fireball 
in $AA$ collisions from the initial entropy density determined 
via the charged particle multiplicity pseudorapidity density,
$dN_{ch}^{AA}/d\eta$,
at mid-rapidity  
($\eta=0$)
calculated from the two component Glauber wounded nucleon
model \cite{KN} 
\beq
\frac{dN_{ch}^{AA}}{d\eta}=\left[\frac{(1-\alpha)}{2}
N_{part}+
\alpha N_{coll}\right]
\frac{dN_{ch}^{pp}}{d\eta}\,,
\label{eq:80}
\eeq
where $dN_{ch}^{pp}/d\eta$ is the multiplicity density for $pp$ collisions,
$N_{part}$ and $N_{coll}$ for a given impact parameter $\bb$ 
are given by the Glauber formulas 
\bea
N_{part}(\bb)=2\int d\ro T_{A}(\ro)\left\{1\right.
\nonumber\\
-
\left.\exp[-T_A(\bb-\ro)\sigma_{in}^{NN}]\right\},
\label{eq:81}
\eea
\beq
N_{coll}(\bb)=\sigma_{in}^{NN}\int d\ro T_A(\ro)T_A(\bb-\ro)\,
\label{eq:82}
\eeq
with $T_A(\bb)=\int dz\rho_A(\bb,z)$ the nuclear
profile function.
The $N_{part}$ and $N_{coll}$ 
have been calculated with the Woods-Saxon nuclear distribution
\beq
\rho_{A}(r)=\frac{\rho_0}{1+\exp[(r-R_A)/a]}
\label{eq:90}
\eeq
with $R_{A}=(1.12A^{1/3}-0.86/A^{1/3})$, and $a=0.54$ fm \cite{GLISS2}.
We use  $dN_{ch}^{pp}/d\eta=2.65$   and $\sigma_{pp}^{in}=35$ mb 
obtained by the UA1 Collaboration \cite{UA1_pp} for non-single-diffractive
events for $pp$ collisions at $\sqrt{s}=200$ GeV.
We take for the fraction of the binary collisions $\alpha=0.135$ 
adjusted to reproduce the experimental STAR data \cite{STAR_NCH} on centrality 
dependence of mid-rapidity $dN_{ch}^{AA}/d\eta$ in Au+Au collisions at
$\sqrt{s}=200$ GeV \cite{Z_MCGL}. 
We determine the entropy density with the help of the 
Bjorken relation \cite{Bjorken}
\beq
s_{0}=\frac{C}{\tau_{0}\pi S_{f}}\frac{dN_{ch}^{AA}}{d\eta}\,.
\label{eq:100}
\eeq
Here $C=dS/dy{\Big/}dN_{ch}^{AA}/d\eta\approx 7.67$ \cite{BM-entropy} 
is the entropy/multiplicity ratio,
and $S_{f}$ is the transverse area of the QGP fireball. 
We define it as the overlapping area of the colliding nuclei.
In calculating the $S_f$ we use the nuclear matter disk radius $r=R_A+ka$
with $k=1.5$ used in our analysis \cite{RAA13}  of the nuclear modification
factor $R_{AA}$.
In the physically reasonable range $k\sim 1-2$,
the results for $I_{AA}$ have a weak dependence on $k$. This
 occurs because for the parton energy loss the change in 
the jet path length with variation of $k$ is approximately 
compensated by the corresponding change in the fireball density.
For Au+Au collisions 
at $\sqrt{s}=200$ GeV 
for $0-12$\% centrality bin as in Ref. \cite{STAR1}
this procedure
gives $T_{0}\approx 327$ MeV
(we take $N_{f}=2.5$ to account for the mass suppression for the strange
quarks in the QGP).
For the above value of the initial temperature in $1+1D$ Bjorken's expansion
the QGP reaches $T\sim T_c$ (here $T_c\approx 160$ is the crossover
temperature) at $\tau_{QGP}\sim 4-5$ fm. We treat the crossover region
as a mixed phase \cite{Bjorken}. For the relevant values of the proper time 
$\tau\lsim 8-10$ fm (see below) 
the QGP fraction in the mixed phase is approximately $\propto 1/\tau$
\cite{Bjorken}. For this reason we can use
in calculating  jet quenching the $1/\tau$ dependence of the number
density of the scattering centers in the whole range of $\tau$
(but with the Debye mass defined for $T\approx T_{c}$ at $\tau> \tau_{QGP}$).

The Bjorken $1+1D$ model \cite{Bjorken}), used in our analysis, neglects
the transverse expansion of the matter that becomes 
very important at $\tau \sim R_A\sim 6$ fm.
However, from the point of view of the parton energy loss its effect should not 
be large due to compensation between the enhancement 
of the energy loss caused by increase of the medium size and its 
suppression caused by reduction of the medium density.
The fact that the transverse expansion of the fireball
does not affect strongly jet quenching in $AA$ collisions
was demonstrated, for the first time, in Ref.~\cite{BMS_hydro}
within the BDMPS approach \cite{BDMPS}.

\subsection{Calculation of $I_{AA}$}
In leading order (LO) pQCD the energy of the hard parton
produced in the direction opposite to the tagged direct photon, $E_T$,
coincides with the photon energy  $E_{T}^{\gamma}$. For
$E_{T}=E_{T}^{\gamma}$ the medium modification
factor for a given photon energy reads
\beq
I_{AA}(z,E_{T}^{\gamma})=\frac{D_{h}^{AA}(z,E_{T}^{\gamma})}
{D_{h}^{pp}(z,E_{T}^{\gamma})}\,,
\label{eq:110}
\eeq
where the nominator and the denominator are 
the FFs for the photon-tagged jets in $AA$ and $pp$ collisions,
respectively. For $pp$ collisions 
\beq
D_{h}^{pp}(z,E_{T}^{\gamma})\!=\!
\!\sum_{i} r_{i}^{pp}(E_{T}^{\gamma})D_{h/i}(z,E_{T}^{\gamma})\,,
\label{eq:120}
\eeq
where
$D_{h/i}$ is the FF for $i\to h$ process
defined by Eq.~(\ref{eq:30})   with $Q=E_{T}^{\gamma}$, and 
$r_{i}^{pp}$ is the fraction of the $\gamma+i$ parton state in the 
$\gamma+$jet events in $pp$ collisions.
The FF for $AA$ collisions
can be written as
\beq
D_{h}^{AA}(z,E_{T}^{\gamma})\!=\!\big\langle\!\big\langle
\!\sum_{i} r_{i}^{AA}(E_{T}^{\gamma})D_{h/i}^{m}(z,E_{T}^{\gamma})
\!\big\rangle\!\big\rangle,
\label{eq:130}
\eeq
where
$D_{h/i}^{m}$ is the medium modified FF for $i\to h$ process
defined by Eqs.~(\ref{eq:20}), (\ref{eq:21})   with $Q=E_{T}^{\gamma}$, and 
$r_{i}^{AA}$ is the fraction of the $\gamma+i$ parton state in the 
$\gamma+$jet events for $AA$ collisions, $\langle\!\langle\,\,\rangle\!\rangle$ means averaging
over the impact parameter of $AA$ collision, 
the jet direction and the position of jet production.
The distribution of the jet production points in the transverse
plane for a given impact vector $\bb$ is  
$\propto T_A(\ro)T_A(\bb-\ro)$.
This averaging procedure 
leads to fluctuations of the fast parton path length
$L$ in the QGP. 
In Fig.~1 we show the distribution in $L$ for Au+Au collision
for the $0-12$\% centrality bin corresponding to the conditions
of the STAR experiment \cite{STAR1} calculated for our model of the QGP
fireball.

\begin{figure} [t]
\vspace{.7cm}
\begin{center}
\epsfig{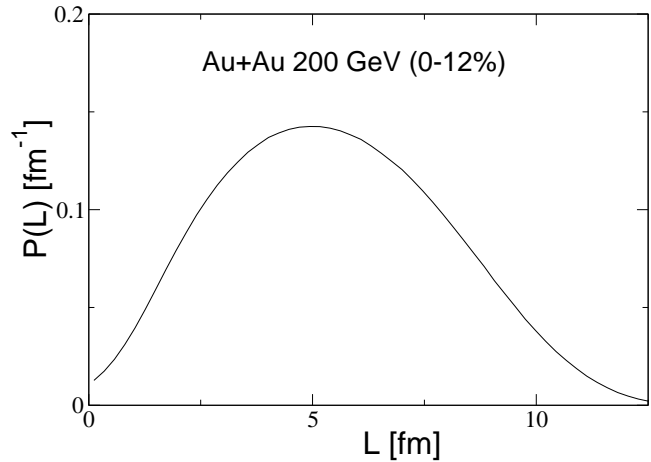}
\end{center}
\caption[.]
{Probability distribution $P(L)$ for the jet path length in the QGP
in our model of the QGP fireball for $0-12$\% centrality bin 
for Au+Au collisions at $\sqrt{s}=200$ GeV.}
\end{figure}

For the STAR experiment \cite{STAR1} $E_{T}^{\gamma}\lsim 20$ GeV. In
this case  the sum over
all relevant types of partons on the right hand side of 
Eqs.~(\ref{eq:120}),  (\ref{eq:130}) 
is dominated by gluon and light quarks. 
Since the medium effects for all light quarks are very similar, 
we consider all light quarks as one effective light quark state  $q$ with
$r_{q}=1-r_{g}$. 
We calculate the $r_{q,g}$ using the LO pQCD 
with the CTEQ6 \cite{CTEQ6} parton distribution functions.
As in the PYTHIA event generator \cite{PYTHIA},
to simulate the higher order effects in  calculation of the partonic
cross sections we take for the virtuality scale in $\alpha_{s}$ the value 
$cQ$ with $c=0.265$. 
For $AA$ collisions we account for the nuclear modification of 
the parton densities
(which leads to a small deviation of $I_{AA}$ from unity even without
parton energy loss) 
with the help of the EKS98 correction \cite{EKS98}.
This calculation gives for $E_T^{\gamma}\sim 12-20$ GeV 
$r_{g}^{pp}/r_{q}^{pp}\sim 0.27-0.33$,  and somewhat smaller values
for $AA$ collisions 
$r_{g}^{AA}/r_{q}^{AA}\sim 0.24-0.3$. 

The LO pQCD formulas do not account for the effect of the 
intrinsic parton transverse momentum on the photon-tagged jet FFs.
The intrinsic parton transverse momenta lead to fluctuation
of the jet energy $E_T$ around the photon energy 
$E_{T}^{\gamma}$ both for $pp$ and $AA$ collisions \cite{Wang1}.
In the pQCD the intrinsic parton transverse momenta in PDFs
emerge in NLO calculations due to the initial state radiation 
in jet production.
The NLO calculations performed in Ref.~\cite{Wang_NLO2} show that for $AA$ collisions
the smearing  correction,
$\Delta_{sm}$, to the medium modification factor $I_{AA}(z)$
at $E_{T}^{\gamma}\sim 7-8$  
blows up at $z\gsim 0.8-0.9$. 
In Appendix B we demonstrate that 
$\Delta_{sm}\approx F(z,E_{T}^{\gamma})dI_{AA}/dz/E_{T}^{\gamma\,\,2}$,
where $F(z,E_{T}^{\gamma})$ is a smooth function of $E_{T}^{\gamma}$.
Using this formula with the help of the results of Ref.~\cite{Wang_NLO2}  
we can obtain the smearing correction to $I_{AA}$ for the conditions
of the STAR experiment \cite{STAR1}.  
The STAR data are obtained for $z\sim 0.15- 0.8$.
The results of Ref.~\cite{Wang_NLO2} show 
that at $E_{\gamma}\sim 7-9$ GeV, in the potentially dangerous region $z\sim 0.8$
$\Delta_{sm}$ increases $I_{AA}$ by $\sim 30$\%.
For the photon energy interval $12 <E_T^{\gamma}<20$ GeV studied in 
Ref.~\cite{STAR1} 
from the formula for $\Delta_{sm}$  we conclude that the smearing
correction to $I_{AA}$ should be $\lsim 8$\% at $z\sim 0.8$
(in fact even at $z=0.9$ it is relatively small),
and for $z\lsim 0.6$ it should be very small. 
For this reason for the photon energy region studied in Ref.~\cite{STAR1}
accuracy of the LO pQCD predictions for $I_{AA}$  should be quite good.

\section{Numerical results and discussion}
The data of Ref. \cite{STAR1} are obtained for the photon energy region 
$12<E_{T}^{\gamma}<20$ GeV. We define the nuclear modification factor 
$I_{AA}$ for the photon energy window $E_1<E<E_2$ (hereafter, 
for notational simplicity, we omit the $T$ and
$\gamma$ indeces of the photon energy)
as
\beq
I_{AA}(z,E_1,E_2)=\frac{D_{h}^{AA}(z,E_1,E_2)}
{D_{h}^{pp}(z,E_1,E_2)}\,,
\label{eq:140}
\eeq
where the FFs in the denominator and numerator
are averaged over the photon energy 
with the help of the LO pQCD 
cross section for photon-jet events $d\sigma/dydE$ 
defined as 
\beq
D_h^{i}(z,E_1,E_1)=\frac{\int_{E_{1}}^{E_{2}}dE D_{h}^{i}(z,E)\frac{d\sigma^i}{dydE}}
{\int_{E_{2}}^{E_{2}}dE \frac{d\sigma^i}{dydE}}\,
\label{eq:150}
\eeq
with superscript $i=AA,pp$. The cross section 
$d\sigma^{AA}/dydE$ is calculated with the EKS98 \cite{EKS98}
correction to the nuclear PDFs.

Before presenting our results for the medium modification 
factor $I_{AA}$, it is instructive to first 
compare our results for the $pp$ photon-tagged FF and to
illustrate 
the $z$ and $L$ dependence of the medium effects in our model.
In Fig.~2 we compare our results for the $pp$ photon-tagged FF
for charged hadrons
for the STAR energy window between $E_1=12$ and $E_2=20$ GeV 
with that measured in Ref.~\cite{STAR1}. 
From Fig.~2 one sees that the theoretical photon-tagged FF agrees
reasonably with that from STAR \cite{STAR1}. 
In Fig.~2 in addition to the prediction for the photon-tagged
jet FF calculated using FFs given by Eq.~(\ref{eq:30})
we also plotted the curve obtained using the pure KKP FFs $D_{h/i}(z,Q)$.
One can see that the results for both the methods agree reasonably well. 
However, in principle, from the point of view of the theoretical 
predictions for $I_{AA}$ the quality of description of the  
$pp$ photon-tagged FF is not very important.

To illustrate 
the relative contribution from gluon and quarks to the photon-tagged FFs 
in Fig.~3 we plot fraction of the gluon contribution to $D_{h}^{pp,AA}$.
The results for $D_{h}^{AA}$ are calculated 
for $0-12$\% central Au+Au collisions (as we noted in Sec.~2 it corresponds
in our model of the QGP fireball 
to the initial QGP temperature $T_0\approx 327$ MeV at $\tau_0=0.5$ fm).
We used $\alpha_s^{fr}=0.5$ that gives a reasonable description
of $I_{AA}$ (see below).
To illustrate better
the medium effects we present predictions for 
$D_{h}^{AA}$ obtained with and without the EKS98 corrections to the nuclear PDFs.
One can see that both for $pp$ and $AA$ collisions the relative 
gluon contribution decreases strongly with increasing $z$. 
The effect of the nuclear modification of the PDFs
is relatively small. From the curves for the Au+Au collision,
one can see that the suppression of the gluon fraction 
at large $z$ is considerably stronger than for
$pp$ collisions. This is a consequence of a bigger energy loss
in the QGP for gluons. 

In order to better demonstrate the difference between the strength of 
jet quenching for gluon and quark jets in Fig.~4 we show the 
$z$-dependence
of the average jet path length in the QGP defined via the medium
modified FFs (\ref{eq:20}) calculated for $\alpha_s^{fr}=0.5$  
for the jet energy
$E_T=15$ GeV  (we included
the argument $L$ that has been omitted for clarity in (\ref{eq:20})) 
\beq
\langle L_i(z)\rangle=
\frac{\int dL L P(L)D_{h/i}^{m}(z,E_T,L)}
{\int dL P(L)D_{h/i}^{m}(z,E_T,L)}\,,
\label{eq:151}
\eeq
where $P(L)$ is the jet path length distribution 
for Au+Au collisions for $0-12$\% centrality bin
shown in Fig.~1. One sees that the typical path 
lengths for gluon and quark jets, that contribute to hadron production
at small $z$, are close to each other. But for large $z$ the typical
jet path length for gluon
jets becomes by a factor of $\sim 1.5-2$ smaller than that for 
quark jets. 
Both for quarks and gluons the value of $\langle L(z)\rangle$
decreases with increase of $z$. It occurs because hadron production 
at large $z$ is biased to events with smaller parton energy loss, i.e. to
events with smaller parton path length through the medium.
The effect is more pronounced for gluons that have larger energy loss.

\begin{figure} [t]
\vspace{.7cm}
\begin{center}
\epsfig{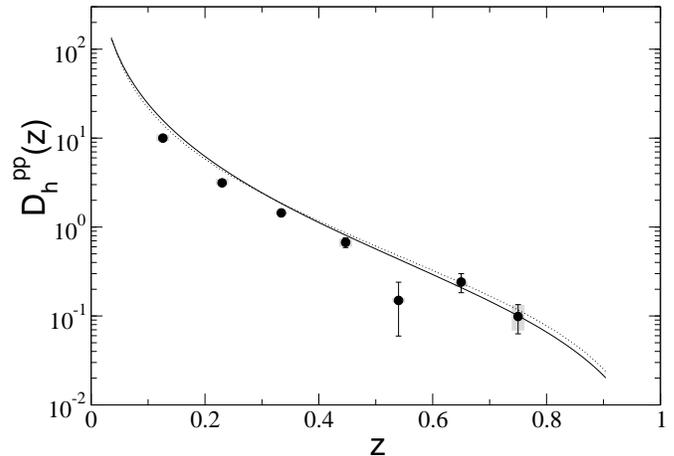}
\end{center}
\caption[.]
{The photon-tagged
jet FF for $12<E_{T}^{\gamma}<20$ GeV for $pp$ collisions
at $\sqrt{s}=200$ GeV calculated using FFs given by Eq.~(\ref{eq:30})
(solid) and obtained for pure KKP FFs $D_{h/i}(z,Q)$ (dotted). 
Data points are from Ref.~\cite{STAR1}.
}
\end{figure}
\begin{figure} [t]
\vspace{.7cm}
\begin{center}
\epsfig{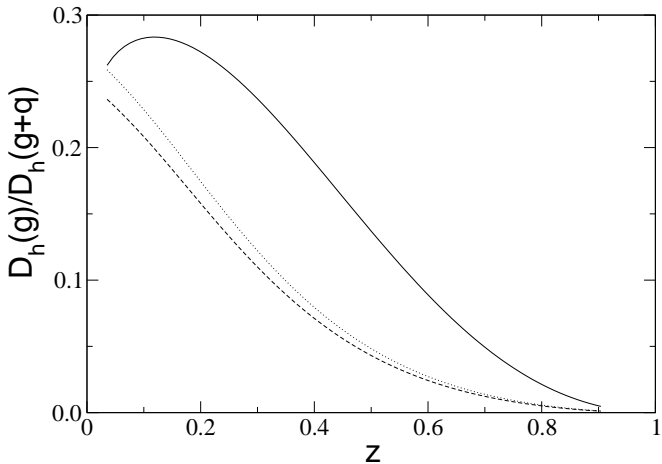}
\end{center}
\caption[.]
{Fraction of the gluon contribution to the photon-tagged
jet FF for $12<E_{T}^{\gamma}<20$ GeV at $\sqrt{s}=200$ GeV
for $pp$ collisions and for Au+Au collisions for $0-12$\% centrality bin.
Solid line: $pp$ collisions; 
dashed and dotted line: Au+Au collisions,
calculations  
for $\alpha_s^{fr}=0.5$ with and without the EKS98 \cite{EKS98} correction.
}
\end{figure}
\begin{figure} [t]
\vspace{.7cm}
\begin{center}
\epsfig{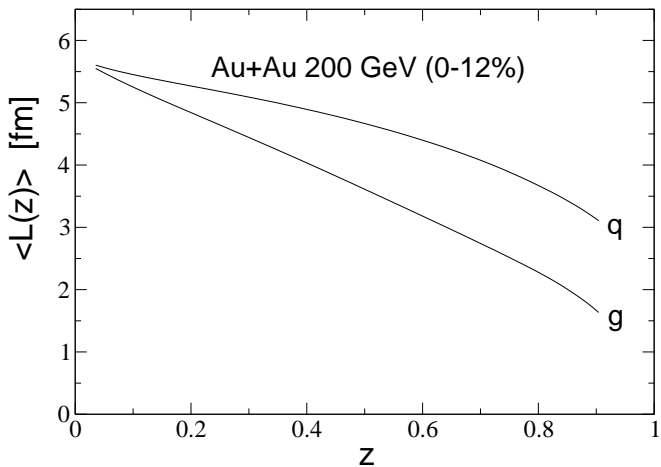}
\end{center}
\caption[.]
{Average jet path length defined by Eq.~(\ref{eq:151}) 
for quark and gluon jets (top to bottom) 
vs $z$ for jet energy $E_T=15$ GeV calculated for $\alpha_s^{fr}=0.5$
with the path length distribution $P(L)$ for 
$0-12$\% centrality bin of Au+Au collisions at $\sqrt{s}=200$ GeV.
}
\end{figure}
\begin{figure} [t]
\vspace{.7cm}
\begin{center}
  \epsfig{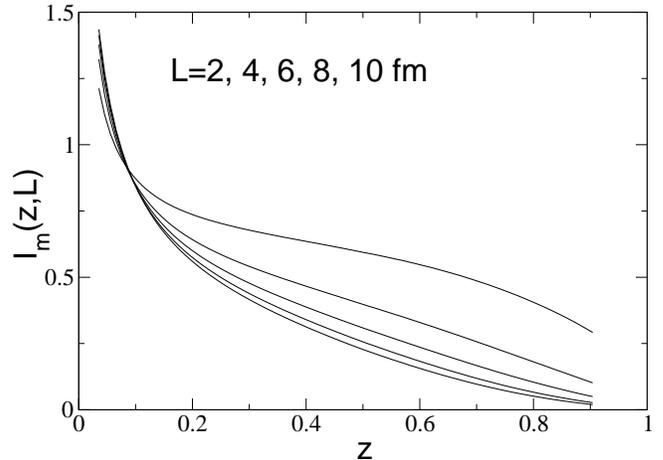}
\end{center}
\caption[.]
{Medium modification factor $I_{m}(z,L)$ for the photon-tagged
jet FFs (\ref{eq:120}), (\ref{eq:160}) averaged over energy in the window $12<E_{T}^{\gamma}<20$ GeV
(see text for details) calculated with $\alpha_s^{fr}=0.5$
for $L=2$, $4$, $6$, $8$, and $10$ fm (top to bottom at large $z$).
}
\end{figure}

To illustrate the $L$-dependence of the medium modification
of the photon-tagged FF, in Fig.~5 we plot the medium
modification factor  for several path lengths in the QGP
(we denote it by $I_{m}(z,L)$).
To separate the medium effect from influence of the nuclear 
corrections to the factors $r_{q,g}$, we calculated $I_m$ 
without the EKS98 corrections, i.e. using for
the nominator the medium-modified 
FFs (\ref{eq:20}) weighted by the factors $r_{q,g}^{pp}$
\beq
D_{h}^{m}(z,E_{T}^{\gamma})\!=\!
\!\sum_{i} r_{i}^{pp}(E_{T}^{\gamma})D_{h/i}^{m}(z,E_{T}^{\gamma})\,,
\label{eq:160}
\eeq
as for $pp$ case (\ref{eq:120}). We perform averaging over energy as 
for the factor $I_{AA}$ given by Eqs.~(\ref{eq:140}), (\ref{eq:150}).
From Fig.~5 one sees that the $L$-dependence of the medium modification
becomes rather weak at $L\sim 8-10$ fm. One can see that, in agreement with
the decrease of $\langle L_{q,g}(z)\rangle$ with increase of $z$ 
seen from Fig.~4, the
photon-tagged FF at large $z$ should be biased to the events
with smaller jet path lengths through the QGP.

Finally, after presenting illustrative results, 
in Fig.~6 we confront our results for the medium modification factor
$I_{AA}$ obtained for $\alpha_s^{fr}=0.4$, $0.5$ and $0.6$
with the data from STAR \cite{STAR1}.
Our previous analyses \cite{RAA13,Z_RPP} of the RHIC data on 
the nuclear modification factor
$R_{AA}$ in Au+Au collisions at $\sqrt{s}=200$ GeV  at $p_T\sim 10-20$ GeV
support $\alpha_{s}^{fr}\sim 0.5$.
From Fig.~6 one sees that $\alpha_{s}^{fr}\sim 0.5$ gives also 
a reasonable agreement with the STAR data \cite{STAR1} on the 
nuclear modification
factor $I_{AA}$ in the whole range of $z$ studied in Ref.~\cite{STAR1}
from $z\sim 0.8$ down to $z\sim 0.15-0.25$.
Note that at $z\lsim 0.4$ our results depend weakly 
on the value of $\alpha_{s}^{fr}$. 

In Fig.~6 we included the region of very small $z$ down to $z\sim 0.03$
that corresponds to hadrons with momentum $\sim 0.4-0.5$ GeV.
It is done just to illustrate the flow of the jet momentum into the soft region
in our model, which satisfies the momentum sum rule.
Of course, the production of such low energy hadrons cannot
be treated in the jet fragmentation picture, because 
it should involve fragmentation of partons with energy comparable
to the energy of the thermal partons in the produced QGP.  
\begin{figure} [t]
\vspace{.7cm}
\begin{center}
  \epsfig{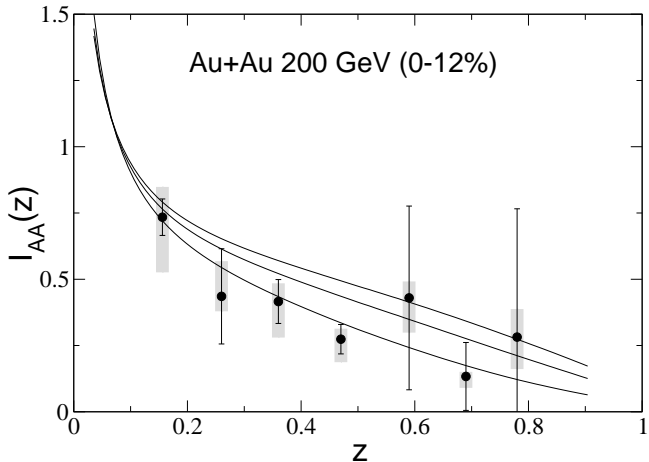}
\end{center}
\caption[.]
{Medium modification factor $I_{AA}$ for the photon-tagged
jet FF  
at $12<E_{T}^{\gamma}<20$ GeV
for Au+Au collisions 
at $\sqrt{s}=200$ GeV for $0-12$\% centrality bin.
The curves are for $\alpha_s^{fr}=0.4$, $0.5$ and $0.6$
(top to bottom at large $z$).
Data points are from Ref.~\cite{STAR1}.
}
\end{figure}

As we noted in Introduction, the approximation of 
independent gluon emission, based on Landau's method
\cite{RAA_BDMS}, has no a rigorous theoretical justification  at $z\ll 1$,
where cascading of the primary gluons radiated from fast partons
may come into play. 
The typical energy for partons fragmenting to hadrons in 
the region $z\sim 0.15-0.25$
is $\sim 3-6$ GeV
(we take the jet energy $E\sim 15$ GeV
that is approximately the mean energy for the STAR window
$E\sim 12-20$ GeV \cite{STAR1}).
It corresponds to the parton fractional momentum $x\sim 0.2-0.4$
The formation length for the induced radiation, $L_f^{in}$, 
of a primary gluon with the fractional momentum $x$ from a fast quark 
can be estimated as \cite{LCPI}
\beq
L_f^{in}\sim\frac{2Ex(1-x)S_{LPM}}{\epsilon^2}\,,
\label{eq:170}
\eeq
where as in Eq.~(\ref{eq:40}) $\epsilon^2=m_q^2x^2+m_g^2(1-x)$, 
$S_{LPM}$ is the LPM suppression factor.
From the formula (\ref{eq:170}) 
one can obtain
$L_f^{in}\sim 3-4$ fm. 
We used $S_{LPM}\sim 0.3$ that corresponds to 
$\tau\sim 3$ fm that is interesting to us. 
The formation length for radiation of the secondary
gluons with energy about half of that for the primary gluons
is about $\sim 2-3$ fm. It means that typically
splitting of the primary gluons can occur at $\tau\sim 5-6$ fm.
In this region the density of the QGP is at least by a factor
of $\sim 10$ smaller than that at the initial time $\tau_0$.
For this reason the induced gluon splitting of the primary radiated
gluons, that are neglected in our analysis, may be a relatively weak effect.

The present analysis assumes that the medium effects are 
present only in $AA$ collisions. It is possible that 
a small-size QGP is produced in $pp$ collisions as well.
The idea that the QGP may be produced in hadron collisions
is very old \cite{Shuryak}.
Application of the 
Bjorken relation (\ref{eq:100}) to $pp$ collisions 
shows that at RHIC energy $\sqrt{s}=200$ GeV the initial temperature
of the mini-QGP fireball may be as large as $\sim 200-230$ MeV
\cite{Z_RPP} 
which is well above the deconfinement temperature.
In the scenario with the mini-QGP production the theoretical 
nuclear modification factor $R_{AA}$ should be divided by
the medium modification factor for $pp$ collisions $R_{pp}$ 
\cite{Z_RPP}
that accounts for jet quenching in the mini-QGP produced 
in $pp$ collisions. For RHIC energy $\sqrt{s}=200$ GeV numerical calculations 
within the LCPI approach give $R_{pp}\sim 0.7-0.8$ at $p_T\sim 10-20$ GeV
\cite{Z_RPP}. The results of Ref.~\cite{Z_RPP} show that in the scenario
with mini-QGP production some increase
of the nuclear modification factor $R_{AA}$ due to the additional
factor $1/R_{pp}$ can be imitated by reduction of the $\alpha_{s}^{fr}$.
Since a direct measurements of $R_{pp}$ is impossible, 
it is practically impossible to distinguish
the scenarios with and without the mini-QGP production in 
$pp$ collisions using the data on the $R_{AA}$.
For the nuclear modification factor $I_{AA}$ the situation
is similar. In the scenario with mini-QGP production in $pp$ 
collisions the theoretical $I_{AA}$ should be divided by the medium
modification factor $I_{pp}$ for $pp$ collisions, which, similarly to
$R_{pp}$, is unobservable quantity. And its effect on the theoretical
$I_{AA}$ can be imitated by some change in  $\alpha_s$.
However, contrary to the medium effects in $pp$ collisions on the high-$p_T$
spectra, the photon-tagged jet FF, at least in principle, enables us 
\cite{Z_pp}
to study the effect of mini-QGP by measuring its variation
with the multiplicity of soft off-jet particles 
(the so-called underlying events,
see Ref.~\cite{Field} for a review).
But this requires measurements of the photon-tagged jet FF
for very high 
underlying event multiplicities $dN_{ch}^{UE}/d\eta\sim 40$ \cite{Z_pp}.
Such multiplicities  are too high to be measured at RHIC energies
in the $\gamma+$jet events.

\section{Summary}
We have calculated the nuclear modification factor $I_{AA}$ 
for the photon-tagged jets within the jet quenching scheme
based on the LCPI approach \cite{LCPI} to the induced radiation
with the collisional energy loss treated as a perturbation
to the radiative mechanism.
The calculations are performed for running $\alpha_s$ frozen
at low momenta. 
Our scheme for calculation of the medium-modified photon-tagged jet 
FF function in $AA$ 
collisions preserves the flavor  and the momentum conservation.
We
have compared the theoretical predictions
with the recent data from STAR \cite{STAR1} for Au+Au collisions
at $\sqrt{s}=200$ GeV for $12 <E_T^{\gamma} <20$ GeV.
We obtained a reasonable agreement with the STAR data
in the whole range of $z$ from $z\sim 0.8$ down to
$z\sim 0.15$ for running coupling constant frozen at the value 
$\alpha_s^{fr}\sim 0.5$. This value agrees well with that
necessary for description of the RHIC data on the nuclear modification factor
$R_{AA}$.

\begin{acknowledgments}
I would like to thank Wenchang Xiang for the invitation to visit Guizhou
University of Finance and Economics, where this work was completed.
This work is supported 
in part by the 
grant RFBR
15-02-00668-a.
\end{acknowledgments}

\renewcommand{\theequation}{A\arabic{equation}}
\setcounter{equation}{0}
\section*{Appendix A. Formulas for 
the spectrum of the induced $a\to bc$ transition }
In this appendix for the reader convenience and completeness we
give formulas for calculations of the probabilities
of $q\to gq$, $g\to gg$, and $g\to q\bar{q}$  transitions in
the LCPI approach \cite{LCPI}.
We use the representation of the induced spectrum 
obtained in Ref.~\cite{Z04_RAA} which is convenient for numerical calculations. 
In general in the LCPI approach 
the $x$-distribution for the induced $a\to bc$ 
partonic transition (hereafter $x=E_b/E_a$) for parton $a$
with momentum along the $z$-axis 
produced
in a hard process at $z=0$ in the matter of thickness $L$  
can be written as
\beq
\frac{d P}{d
x}=
\int\limits_{0}^{L}\! d z\,
n(z)
\frac{d
\sigma_{eff}^{BH}(x,z)}{dx}\,,
\label{eq:a10}
\eeq
where $n(z)$ is the medium number density, $d\sigma^{BH}_{eff}/dx$ 
is an effective Bethe-Heitler
cross section. It  accounts for both the LPM and finite-size effects
and can be written as
\bea
\frac{d
\sigma_{eff}^{BH}(x,z)}{dx}=-\frac{P_{a}^{b}(x)}
{\pi M}\mbox{Im}
\int\limits_{0}^{z} d\xi \alpha_{s}(Q^{2}(\xi))
\nonumber\\
\times
\left.
\exp{\left(-i\frac{\xi}{L_f}\right)}
\frac{\partial }{\partial \rho}\left(\frac{F(\xi,\rho)}{\sqrt{\rho}}\right)
\right|_{\rho=0}\,\,.
\label{eq:a20}
\eea
Here 
$P_{a}^{b}(x)$
is the usual pQCD splitting
function for $a\to bc$ transition,
$
M=E_ax(1-x)\,
$,
$L_{f}=2M/\epsilon^{2}$
with 
$\epsilon^{2}=m_{b}^2(1-x)+m_c^2x-m_{a}^{2}x(1-x)$,
$Q^{2}(\xi)=aM/\xi$ with $a\approx 1.85$ 
(this value of the parameter $a$ was fixed in
Ref.~\cite{Z_Ecoll} by comparison of the induced gluon
spectrum calculated in the coordinate representation with that
obtained in the momentum representation for the dominant $N=1$ rescattering
contribution),
$F$ is the solution to the radial Schr\"odinger 
equation for the azimuthal quantum number $m=1$ 
\bea
i\frac{\partial F(\xi,\rho)}{\partial \xi}=
\left[-\frac{1}{2M}\left(\frac{\partial}{\partial \rho}\right)^{2}
+v(\rho,x,z-\xi)\right.
\nonumber\\
+\left.\frac{4m^{2}-1}{8M\rho^{2}}
\right]F(\xi,\rho)\,
\label{eq:a30}
\eea
with the potential 
\beq
v(\rho,x,z)=-i\frac{n(z)\sigma_{3}(\rho,x,z)}{2}\,.
\label{eq:a31}
\eeq
Note, that in terms of the original longitudinal variable $z$ along the fast
parton momentum we solve the Schr\"odinger equation backward in time/$z$.
The point $\xi=0$ corresponds to the last rescattering of the $bc\bar{a}$
system on a medium constituent located at $z$. This technical trick
allows us to have a smooth boundary condition for $F$ at $\xi=0$: 
$F(\xi=0,\rho)=\sqrt{\rho}\sigma_{3}(\rho,x,z)
\epsilon K_{1}(\epsilon \rho)$  
($K_{1}$ is the Bessel function). 
The function $\sigma_{3}(\rho,x,z)$ is the cross section of interaction
of the $bc\bar{a}$ system with a medium constituent
located at $z$ (the argument $\rho$ is the transverse
distance between $b$ and $c$). 
In the transverse plane the parton $\bar{a}$ in the
$bc\bar{a}$ system is located at the center of mass of the $bc$ 
pair. 
In the Schr\"odinger equation
(\ref{eq:a30}) $M$ plays the role of the reduced "Schr\"odinger mass"
for the $bc$ pair (since the ``masses'' for the $b$ and $c$
partons are $E_ax$ and $E_a(1-x)$, respectively).

The three-body cross section $\sigma_{3}$ can be written in terms of the well known
dipole cross section for the color singlet $q\bar{q}$ pair \cite{NZ_sigma3}
that is given by
\beq
\sigma_{q\bar{q}}(\rho,z)=C_{T}C_{F}\int d\qb
\alpha_{s}^{2}(q^{2})
\frac{[1-\exp(i\qb\ro)]}{[q^{2}+\mu^{2}_{D}(z)]^{2}}\,.
\label{eq:a50}
\eeq
Here $C_{F,T}$ are the color Casimir for the quark and thermal parton 
(quark or gluon), and $\mu_{D}(z)$ is the local Debye mass.
In terms of the dipole cross section (\ref{eq:a50}) the three-body cross
sections for
$q\to gq$, $g\to gg$, and $g\to q\bar{q}$  transitions
read
\bea
\left.\sigma_{3}(\rho,x,z)\right|_{q\to gq}=\frac{9}{8}
[\sigma_{q\bar{q}}(\rho,z)+
\sigma_{q\bar{q}}((1-x)\rho,z)]
\nonumber\\
-\frac{1}{8}\sigma_{q\bar{q}}(x\rho,z)\,,
\label{eq:a60}
\eea
\bea
\left.\sigma_{3}(\rho,x,z)\right|_{g\to gg}
=\frac{9}{8}
[\sigma_{q\bar{q}}(\rho,z)+
\sigma_{q\bar{q}}((1-x)\rho,z)
\nonumber\\
+\sigma_{q\bar{q}}(x\rho,z)]\,,
\label{eq:a70}
\eea
\bea
\left.\sigma_{3}(\rho,x,z)\right|_{g\to q\bar{q}}=\frac{9}{8}
[\sigma_{q\bar{q}}(x\rho,z)+
\sigma_{q\bar{q}}((1-x)\rho,z)]
\nonumber\\
-\frac{1}{8}\sigma_{q\bar{q}}(\rho,z)\,.
\label{eq:a80}
\eea

The LPM suppression and finite-size suppression 
for the in-medium $a\to bc$ process is characterized by
the factor
\beq
S(M,x)=\frac{dP/dx}{dP^{BH}/dx}\,,
\label{eq:a90}
\eeq
where the denominator is the Bethe-Heitler spectrum 
calculated using in Eq.~(\ref{eq:a10})
the real Bethe-Heitler cross section for $a\to bc$ transition
given by \cite{QQbar}
\beq
\frac{d\sigma^{BH}(x,z)}{dx}=
\int d\ro |\Psi_{a}^{bc}(\ro,x)|^{2}\sigma_{3}(\rho,x,z)\,,
\label{eq:a100}
\eeq
where $\Psi_{a}^{bc}(\ro,x)$ is the light-cone wave function for 
$a\to bc$ transition. Since  $|\Psi_{a}^{bc}(\ro,x)|^{2}$ contains the
splitting function $P_a^b(x)$\footnote{It is worth noting that the formula 
for the Bethe-Heitler cross section (\ref{eq:a100}) exactly corresponds
to that for the effective Bethe-Heitler cross section given by 
Eq.~(\ref{eq:a20}) with the infinite upper limit of the $\xi$-integration
and the function $F$ calculated for $v=0$.},
the factor $S(M,x)$ for a given value of $M$  has a smooth
dependence on $x$. It occurs  because 
the $x$-dependence of the integrand of (\ref{eq:a20}), that comes only
from the $x$-dependence of $\epsilon^2$ and of the
three-body cross section $\sigma_3$, is relatively weak. 
This fact allows one to reduce considerably  CPU  
time in numerical calculations by
creating a grid of values of $S(M,x)$ in the $M-x$ plane with relatively 
small number of points in 
$x$. Then this grid can be used for
calculation of the spectrum $dP/dx$ by performing interpolation in $x$ 
and using the Bethe-Heitler spectrum that does not require 
much computing power.

In the above formulas for the effective Bethe-Heitler cross section
 (\ref{eq:a20}) and for the dipole cross section (\ref{eq:a50}) 
we use the following parametrization for $\alpha_s(Q^2)$ 
\beq
\alpha_s(Q^2) = 
\begin{cases}
\alpha_{s}^{fr} & \text{if } Q \le Q_{fr}\;, \\
\dfrac{4\pi}{9\log(Q^2/\Lambda_{QCD}^2)}  & \text{if } Q > Q_{fr}
\end{cases}
\label{eq:a110}
\eeq
with $Q_{fr}=\Lambda_{QCD}\exp\left(
2\pi/9\alpha_{s}^{fr}\right)$, $\Lambda_{QCD}=300$ MeV.

\renewcommand{\theequation}{B\arabic{equation}}
\setcounter{equation}{0}
\section*{Appendix B. The smearing
correction to
$I_{AA}$.}
In this appendix we discuss the energy dependence of the smearing
correction to the LO predictions for the nuclear modification factor
$I_{AA}$. This correction potentially may be important at
$z$ close to unity.
The question is where the regime of large smearing correction begins.
To understand this we use as a plausible estimate of the smearing effect
the results of the NLO model \cite{Wang_NLO2}. 
This analysis shows that for 
$AA$ collisions the smearing correction to
the medium modification factor $I_{AA}(z)$
blows up at $z\gsim 0.8-0.9$ for $E\sim 8$ GeV. 
The results of Ref.~\cite{Wang_NLO2} can be easily rescaled to our
conditions. Indeed, let us write the LO $I_{AA}$ as
\beq
I_{AA}(z)=D_{h}(z+\Delta z)/D_{h}(z)\,,
\label{eq:b10}
\eeq
where $D_h$ is the FF for $pp$ collisions, 
$\Delta z=z\Delta E/E$ is shift of $z$ due to the radiative parton
energy loss $\Delta E$ in the QGP (here, for simplicity, we omit averaging
over $\Delta z$ which should be made, also for clarity  we omit the 
argument $E$ and superscript $pp$ on the FF).
The ${I}_{AA}$ in the presence of the smearing
(we denote it $\bar {I}_{AA}$)
 can be written
as 
\bea
\bar{I}_{AA}(z)=[D_{h}(z+\Delta z)+D_{h}^{''}(z+\Delta z)
\langle \delta z^{2}\rangle /2]\nonumber\\
\times [D_{h}(z)+D_{h}^{''}(z)\langle \delta z^{2}\rangle/2]^{-1}\,,
\label{eq:b20}
\eea
where $\delta z=z q/E$ and $q$ is the shift of the jet energy
($E_{jet}=E+q$) (we assume that the smearing correction is not very large,
and  keep only the second order terms
in $\delta z$). 
From (\ref{eq:b20})
we obtain
an equation which does not contain $\Delta z$
\beq
\bar{I}_{AA}(z)\approx I_{AA}(z)+\Delta_{sm}\,,
\label{eq:b30}
\eeq
\bea
\Delta_{sm}\approx[2 I_{AA}^{'}(z)D_{h}^{'}(z)+
I_{AA}^{''}(z)D_{h}(z)]\nonumber\\ 
\times z^{2}\langle q^{2}\rangle/2E^{2}\,.
\label{eq:b40}
\eea
The second term in the square brackets in Eq.~(\ref{eq:b40})  
can be neglected (since
$I_{AA}$ is a smooth function as compared to $D_{h}$).
Then we obtain
\beq
\Delta_{sm}\approx F(z,E)I_{AA}^{'}(z)/E^{2}\,,
\label{eq:b50}
\eeq
where $F(z,E)$ is a smooth function of energy, and does not depend on the
strength of the medium suppression at all.
The fact that $\Delta_{sm}\propto dI_{AA}/dz$ is quite natural. For a 
flat $I_{AA}$ the smearing effect should vanish
since the numerator and denominator are affected in the same way.
Note that the $\Delta_{sm}\propto dI_{AA}/dz$ scaling agrees with
the results for $\Delta_{sm}$ from Ref.~\cite{Wang_NLO2} obtained
for different magnitudes of the energy loss (shown in Fig.~2 
of Ref.~\cite{Wang_NLO2}).
So now with the help of (\ref{eq:b50})
we can rescale the smearing correction
of Ref.~\cite{Wang_NLO2} to the higher energy region corresponding
to the STAR experiment \cite{STAR1}.

\vskip .5 true cm

\end{document}